# SPDC in ppLN ridge waveguide: an analysis for efficient twin photon generation at 1550 nm.

RAMESH KUMAR,[1]  JOYEE GHOSH[1,*]

[1]Department of physics, Indian Institute of Technology Delhi, New Delhi 110016
*E-Mail: _joyee@physics.iitd.ac.in_

Abstract. We study modal characteristics of a customized ridge waveguide in Lithium Niobate designed to generate twin photons at telecom wavelength. A quantum analysis of SPDC predicts the possible down conversion processes and optimizes the input beam parameters for fundamental mode emission. Further, a Joint Spectral Amplitude (*JSA*) analysis ensures the generation of signal/idler photons at 1550 nm in the customized LN waveguide. A calculation of the parametric down conversion (PDC) signal power shows a dependence of $\mathcal{L}^{3/2}$ for a waveguide compared to a linear dependence in case of a bulk crystal.

## 1. INTRODUCTION

Photon pairs are an important tool for various applications in quantum communication and quantum computing. Spontaneous parametric down-conversion (SPDC), utilizing the second-order nonlinearity: $\chi^{(2)}$ of the medium, has evolved as a promising technique to generate single or paired photons, due to its high efficiency and simplicity [1]. In this process a single frequency pump photon of specific energy ($\omega_p$) and wave vector ($\boldsymbol{k_p}$) interacts with a non-linear $\chi^{(2)}$ medium to generate two photons of lower frequencies called signal ($\omega_s$, $\boldsymbol{k_s}$) and idler ($\omega_i$, $\boldsymbol{k_i}$). The process takes place under conservation of energy such that, $\omega_p = \omega_s + \omega_i$. Momentum conservation is achieved through quasi-phase matching which is a flexible technique to extend the range of such three-wave mixing processes in periodically poled crystals. SPDC in non-linear bulk crystals suffer low efficiency due to problems such as design complications, beam overlaps, small interaction length, etc. In recent years, there is a lot of attention to SPDC in nonlinear waveguides [2-5], where the optical fields are confined spatially to its transverse cross-section that enhances the degree of nonlinear interaction, and hence its efficiency. Confinement also leads to a generation of photons in well-defined spatial modes that can be efficiently coupled to single-mode optical fibers, which is an advantage in quantum information and communication applications.

In general, waveguides support several spatial modes for the propagating pump, signal and idler photons, corresponding to their respective wavelengths. SPDC involving each set of such three spatial modes leads to a unique spectrum or frequency distribution of the emitted photon pairs. All these distributions superimpose at the output of the waveguide. However, many quantum optics experiments require single photons in a controlled single spatial mode. For such experiments, it is useful to suppress all higher order spatial modes and maximize the photon pair generation in the fundamental mode. In this paper, we present an analysis of the spatial mode structure and the joint biphotonic state that can be emitted through a customized ppLN ridge waveguide, designed to generate twin photons at telecom wavelength. In Section 2, we present the modal analysis of the customized waveguide structure and optimize the input beam parameters for efficient coupling into the same. In Section 3, we simulate and discuss some of down-conversion processes possible through this waveguide. In Section 4, we study various aspects of the joint state of the photon pairs generated in the dominant SPDC process. There are fewer works related to the spectral density and signal power in a waveguide [6, 7]. However, an explicit expression or plot for comparing the generated signal power in case of a waveguide as compared to bulk crystals is missing in the current literature, to the best of our knowledge. For the first time, we have attempted to derive an expression for the signal power expected in case of a waveguide and discussed our finding with respect to bulk crystals. This is detailed in Section 5. We summarize and conclude our results in Section 6.

## 2. MODAL ANALYSIS OF A RIDGE WAVEGUIDE IN LITHIUM NIOBATE

As a medium, Lithium Niobate (LN) is widely used for various nonlinear and integrated applications, because of its transparency over a wide range of frequency (0.35 – 4.5 μm) and high nonlinear coefficient [8]. We analyzed a customized (5% MgO doped) LN ridge waveguide that is periodically poled and type II phase-matched to optimize the conversion process: 775nm → 1550nm. A ridge structure is particularly promising due to its ability to achieve high confinement of light in comparison with other traditional waveguides and guarantees high conversion efficiency over large bandwidth of operation. Also, in comparison to other waveguides, ridge waveguides show particularly high power handling capabilities, which are required for nonlinear optical processes in general [9].

Fig. 1(a) depicts a schematic of this process in the waveguide whose cross section is shown in Fig. 1(b), with D = 2.8 μm, W = 6 μm, H = (5 ± 2) μm. The curved regions at the core boundaries are a consequence of the limitation of the technique used for fabricating this waveguide. The main challenge to fabricate a ridge waveguide comes from the fact that LiNbO$_3$ is a hard and relatively inert material, hence it is relatively difficult to etch this structure. It is also difficult to find suitable masks, with high enough selectivity to allow a deep etching [10,11].

For core and cladding, we modeled the refractive index profile (corresponding to the squared portion in Fig. 1(b)) for the pump beam as demonstrated in Fig. 1(c). Its corresponding fundamental mode supported by this waveguide structure at 775 nm is shown in Fig. 1(d). Using the above model of the waveguide, the guided modes are computed at different wavelengths. Owing to a large cross-section and high index contrast ($\Delta n = 0.8$) this waveguide is found to support many eigen modes. We have used this data for modeling the possible down conversion processes.

We simulated different heights (H) of the waveguide and calculated the effective indices of the TE and TM modes at 775 nm corresponding to the pump and at 1550 nm, corresponding to signal and idler. This is plotted in the Fig. 2, where we find that the effective index

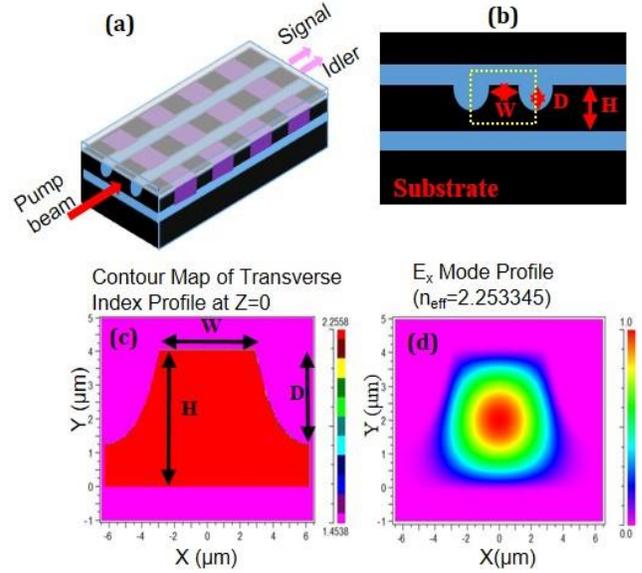

**Figure. 1**: (a) Schematic of PDC in a non-linear waveguide; (b) cross-section of the customized ppLN waveguide; (c) refractive index profile of the modelled ppLN waveguide; (d) fundamental (0, 0) mode profile supported by the wavelength at 775nm.

increases with height and hence the effective interaction area of the waveguide.

We have used the Sellmeier equation [12, 13] for LN to calculate its refractive indices at different wavelengths and polarizations. The group index of a mode is related to the refractive index of the material and is given by:
$n_\xi^g(\lambda) = n_\xi(\lambda) - \lambda \frac{dn_\xi(\lambda)}{d\lambda}$ , where $n_{\xi=(p,s,i)}$ corresponds to refractive index of pump, signal and idler respectively.

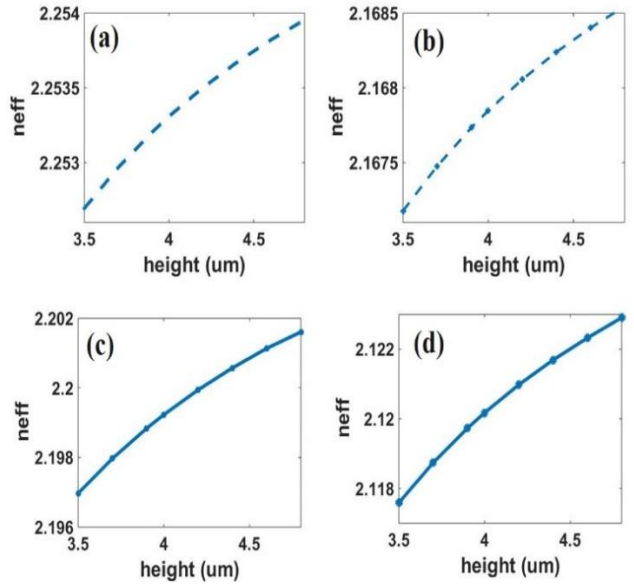

**Figure. 2:** Effective index for the (a) pump mode (TE); (b) pump mode (TM); (c) signal mode; (d) idler mode.

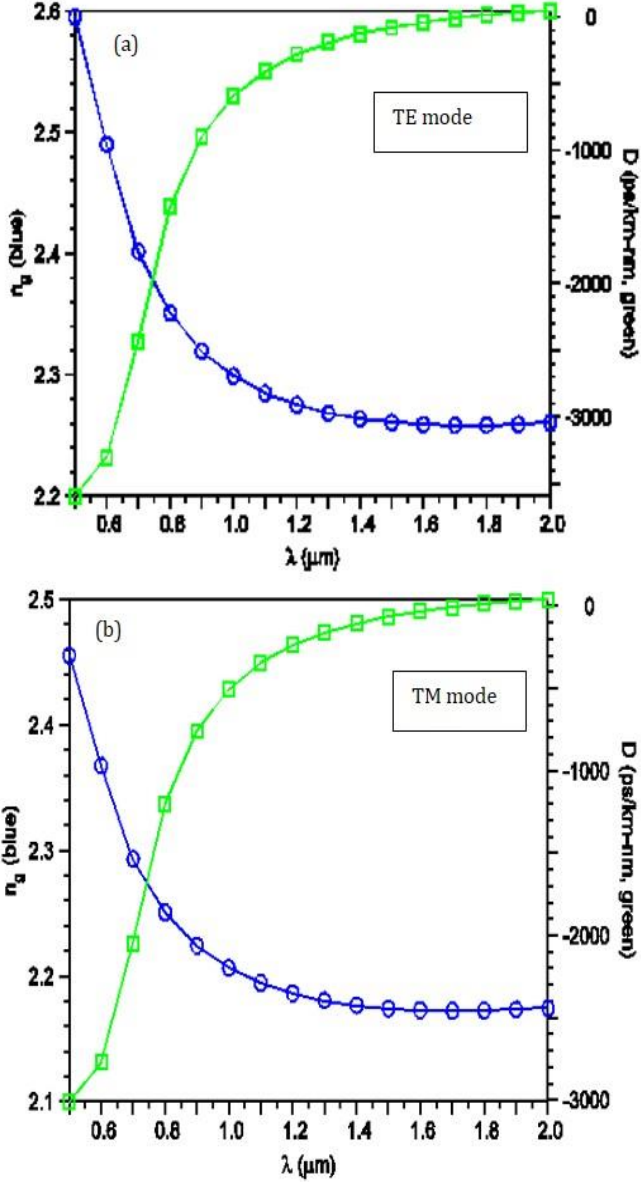

**Figure. 3:** Dispersion (green) and group index (blue) as a function of wavelength (a) TE mode; (b) TM moode.

The total dispersion $D = D_w + D_m$, in a waveguide depends on the waveguide dispersion ($D_w$) and material dispersion ($D_m$), the latter being related to the group velocity dispersion (GVD) as $D_m = -\frac{2\pi c}{\lambda^2} GVD$. For the ppLN ridge waveguide considered, we have investigated the total dispersion and group index as a function of the wavelength, demonstrated in Fig. 3. The results indicate that dispersion at 1550 nm is very low, as desirable for signal and idler photon generation at that wavelength. This therefore enhances the signal power that can be generated at that wavelength, which is further discussed and derived in Sec.6.

**Optimization of Input Beam Parameters**

A majority of the experiments carried out in quantum optics, and especially for quantum information applications, require single photons in a well-controlled single spatial mode. For this, all higher order mode processes need to be suppressed in order to maximize the emission of the photon pairs into the fundamental (0, 0) mode [14]. The pump beam could have a Gaussian profile, obtained by spatial filtering or through a single mode fiber. The launch position of the pump beam into the waveguide dictates the structure of the output state of the down-converted photons. For an optimal Gaussian pump beam, aligned properly, most of its energy goes into the fundamental (0, 0) waveguide mode and higher-order modes are suppressed. Higher-order pump modes also lead to *other* down conversion processes which should be avoided. We examine the overlap integral between the incident free-space Gaussian pump mode and the corresponding modes in the waveguide basis in order to estimate the optimal pump mode. This overlap integral is an indication of the amount of power that can be coupled into a particular pump mode of the waveguide, and is given by

$$A_p^{(l)} = \iint d\mathbf{r}\, u_p^{(l)}(\mathbf{r}) E_{in}^{\text{pump}}(\mathbf{r}). \quad (1)$$

Here $u_p^{(l)}$ is the normalized field profile of a pump mode $l$, supported by the waveguide and the input field $E_{in}^{\text{pump}}(\mathbf{r})$ is the input Gaussian function, defined as:

$$E_{in}^{\text{pump}}(\mathbf{r}) = e^{-\left(\frac{x^2}{a^2}+\frac{y^2}{b^2}\right)}. \quad (2)$$

Where, $2a$ and $2b$ are the width and height of the Gaussian beam cross-section, respectively at which the field amplitude falls to $1/e$ of its axial value.

Table. 1: Optimal input beam parameters

| Parameter | Optimum value ($\mu m$) |
|---|---|
| Launch position (X-direction) | 0 |
| Launch position (Y-direction) | 2.0 |
| Gaussian X-width | 5.5 |
| Gaussian Y-width | 3.5 |

Table. 2: Overlap integrals of various modes at 775 nm

| Mode number | Overlap Integral |
|---|---|
| (0,0) | 0.96 |
| (1,0) | 3.8x $10^{-13}$ |
| (2,0) | 0.0045 |
| (0,1) | 8.55x $10^{-5}$ |

For a given input Gaussian profile various mode profiles of the ppLN waveguide are used to compute the overlap integrals. The amount of power coupled into a particular mode depends on the input beam parameters and the launch position of the beam. Table 1 gives the optimal input beam parameters. With the help of cylindrical optics, it is easy to generate such optimized beam profiles for efficient coupling which could be achieved through mode matching with the waveguide dimension by choosing appropriate lens objectives, for example. In Table 2, the overlap integrals for the first few modes are computed using the input parameters in Table 1. As expected, the overlap integral of the fundamental (0, 0) mode of the waveguide (Fig. 1(d)) with such an input beam is maximum in comparison to that of higher order modes supported by the waveguide.

Also, since the effective index of the waveguide varies with its height 'H' (as seen in Fig.2), we optimized this parameter for generating degenerate signal/idler photons at 1550 nm corresponding to fundamental mode conversion (see (i) in Fig. 4 and Table 3 below). This optimized height was found to be 4.04 μm and the same has been kept for obtaining all subsequent results.

## 3. POSSIBLE DOWN-CONVERSION PROCESSES

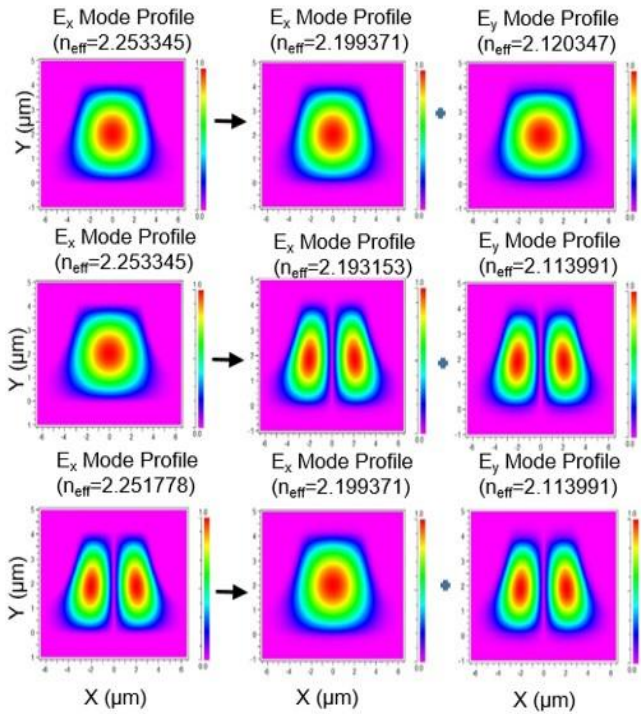

Fig. 4: Mode conversions to different orders of the signal-idler spatial modes in type II PDC process for the given waveguide structure: (i) $(0,0) \rightarrow (0,0) + (0,0)$; (ii) $(0,0) \rightarrow (1,0) + (1,0)$; (iii) $(1,0) \rightarrow (0,0) + (1,0)$.

Due to a large cross-section of the waveguide and a large value of $\Delta n = 0.8$, the structure was highly multi-mode. As a result, different SPDC processes corresponding to generation of different spatial modes are possible. We calculated the different generated wavelengths corresponding to different spatial modes. This was done using the following expression:

$$\frac{2\pi}{\lambda_p}(n_{\text{eff}})_p - \frac{2\pi}{\lambda_s}(n_{\text{eff}})_s - \frac{2\pi}{\lambda_i}(n_{\text{eff}})_i - \frac{2\pi}{\Lambda} = 0. \quad (3)$$

Here $\lambda_{s,i,p}$ are signal, idler and pump wavelengths, respectively.

In Fig. 4 we display the different normalized mode profiles corresponding to different signal and idler wavelengths (elaborated in Table 3) generated in some of the PDC processes possible in the given waveguide structure. The spatial overlap of the three interacting modes $l$, $m$ and $n$ is given as:

$$A_{lmn} = \int_A d\mathbf{r}\, u_p^{(l)}(\mathbf{r})\, u_s^{(m)}(\mathbf{r})\, u_i^{(n)}(\mathbf{r}). \quad (4)$$

We calculated the coupling constants (in terms of these overlap integrals $A_{lmn}$) of the below mode conversion processes to estimate the efficiency in each case. Table 3 gives the computed values of the signal/idler wavelengths and corresponding $A_{lmn}$ for different mode conversions.

Table. 3: Some possible PDC processes generating different wavelengths of signal/idler with their corresponding mode overlap integrals at a fixed pump wavelength of 775 nm.

| $E^{(p)}(l_x, l_y) \rightarrow E^{(s)}(m_x, m_y) + E^{(i)}(n_x, n_y)$ | $\lambda_s$ (nm) | $\lambda_i$ (nm) | $A_{lmn}$ |
|---|---|---|---|
| (i) (0,0) → (0,0) + (0,0) | 1550.0 | 1550.0 | 0.2647 |
| (ii) (0,0) → (1,0) + (1,0) | 1337.5 | 1842.7 | 0.2035 |
| (iii) (1,0) → (0,0) + (1,0) | 1493.6 | 1610.8 | 0.2067 |
| (iv) (0,1) → (0,1) + (0,0) | 1301.3 | 1916.2 | 0.2168 |
| (v) (2,0) → (1,0) + (1,0) | 1460.4 | 1651.3 | 0.1483 |
| (vi) (2,0) → (2,0) + (0,0) | 1388.8 | 1753.5 | 0.1798 |

It shows that the first fundamental mode conversion process: $[(0,0) \rightarrow (0,0) + (0,0)]$ occurs with maximum efficiency corresponding to the highest value of $A_{lmn}$ and corresponds to the degenerate case for the photon pair emission, as expected from the customized structure.

## 4. PDC ANALYSIS FOR A WAVEGUIDE

For SPDC in a waveguide, the general PDC Hamiltonian is given by

$$\hat{H}(t) = \frac{\varepsilon_o}{2}\int d^3\mathbf{q}\, \{\chi^{(2)}:E_p^{(+)}(\mathbf{q},t)E_s^{(-)}(\mathbf{q},t)E_i^{(-)}(\mathbf{q},t) + h.c\}. \quad (5)$$

Where $p$, $s$ and $i$ stand for the pump, signal and idler, respectively $E(\mathbf{q},t)$ is the electric field inside the nonlinear medium at position $\mathbf{q}$ and time $t$. We can derive the normalized guided wave PDC state to be:

$$|\psi\rangle = B\sum_{lmn} A_p^l A_{lmn} \iint d\omega_s d\omega_i\, f_{lmn}(\omega_s,\omega_i)$$
$$\times \hat{a}_s^{(m)\dagger}(\omega_s)\hat{a}_i^{(n)\dagger}(\omega_i)|0,0\rangle. \quad (6)$$

Here, all constants are merged into a dimensionless constant factor $B = \dfrac{d\sqrt{2P_p}}{\mathcal{L}\sqrt{\varepsilon_o c n_s^2 n_i^2 n_p}}$. Here $d$ is the nonlinear coefficient, $P_p$ is the pump power, $\mathcal{L}$ is the length of the waveguide, $n_{s,i,p}$ are the refractive indices of the signal, idler and pump fields. The signal and idler creation operators are written as $\hat{a}_s^{(m)\dagger}$ and $\hat{a}_i^{(n)\dagger}$. We can tune the spread of the photons into different spatial modes by controlling the excitation of the incoming pump wave into the different guided modes $A_p^l$. Thus, if we explicitly consider the spatial modes propagating inside the guided material, it is seen that the generated biphoton state is emitted into a superposition of interacting mode triplets (*lmn*). Due to the overlap and interaction between the pump, signal and idler fields, given by the overlap integral $A_{lmn}$, each such mode triplet exhibits a different overall down-conversion efficiency. In addition, each triplet possesses a unique spectrum $f_{lmn}(\omega_s,\omega_i)$ corresponding to different longitudinal wave vectors satisfying the phase-matching condition. This unique spectrum of the down-converted photon pairs, otherwise known as the two-photon *Joint Spectral Amplitude (JSA)*, is defined as:

$$f_{lmn}(\omega_s,\omega_i) = \alpha(\omega_s+\omega_i)\phi_{lmn}(\omega_s,\omega_i). \quad (7)$$

The *JSA* is a product of two terms. The first term is the *pump envelope function (PEF)*: $\alpha(\omega_s+\omega_i) = e^{-\left(\frac{\omega_s+\omega_i-\omega_p}{\sigma_p}\right)^2}$, which determines the frequencies that satisfy the conservation of energy. The second term is the *phase-matching function (PMF)*, $\phi_{lmn}(\omega_s,\omega_i)$, which depends on the material properties of the non-linear medium used for PDC and geometry of the waveguide [15]. It determines the frequencies that are suitable for the phase matching condition. It is defined as:

$$\phi_{lmn}(\omega_s,\omega_i) = \mathrm{sinc}\left[\Delta\beta_{lmn}(\omega_s,\omega_i)\frac{\mathcal{L}}{2}\right]\exp\left(i\Delta\beta_{lmn}(\omega_s,\omega_i)\frac{\mathcal{L}}{2}\right). \quad (8)$$

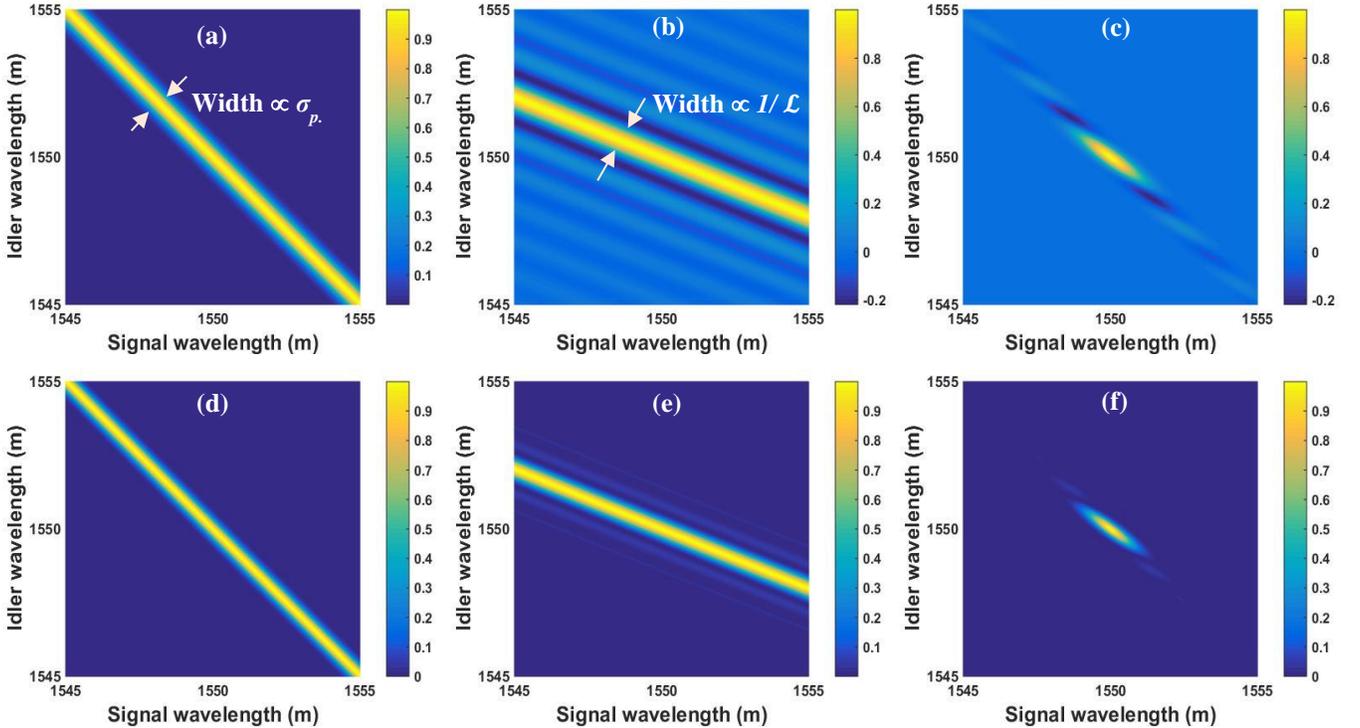

Fig. 5: (a) Pump envelope function. (b) Phase matching function. (c) Joint spectral amplitude function. (d) Pump envelope intensity corresponding to pump envelope function. (e) Phase matching intensity corresponding to phase matching function. (f) Joint spectral intensity corresponding to *JSA*. For all plots we have used pump bandwidth $\sigma_p$=250 GHz, poling period $\Lambda = 8.29\,\mu m$, pump wavelength $\lambda_p = 775\,nm$, length of the waveguide $\mathcal{L} = 1$ cm.

Where $\Delta\beta_{lmn}(\omega_s, \omega_i) = \beta_p^{(l)}(\omega_s + \omega_i) - \beta_s^{(m)}(\omega_s) - \beta_i^{(n)}(\omega_i) - \beta_{QPM}$ is the momentum mismatch between different propagation constants in a waveguide, corrected by the quasi-phase-matching vector, $\beta_{QPM}$. And hence the phase-matching condition would be given by $\Delta\beta_{lmn}(\omega_s, \omega_i) = 0$. Here $\beta_{p,s,i}^{(\tau)} = \frac{2\pi}{\lambda_{p,s,i}} n_{p,s,i}^{(\tau)}$ is the wave vector of the pump, signal and idler, respectively corresponding to the effective indices $n_{p,s,i}^{(\tau)}$ of the same inside the waveguide, where $\tau = (l, m, n)$ modes.

For a pulsed laser source acting as the pump and assuming its spectrum to be a Gaussian distribution with bandwidth ($\sigma_p$ = 250 GHz) in Fig. 5 we have plotted the above features for the customized LN structure for the fundamental mode emission.

In Fig. 5(a) we plot the pump envelope function, $\alpha(\omega_s + \omega_i)$, whose width is directly proportional to the bandwidth $\sigma_p$. The pump envelope function or the energy term varies as a straight line with negative slope that corresponds to a perfect anti-correlation of the two emitted frequencies.

In Fig. 5(b), we have plotted the phase matching function, $\phi_{lmn}$ (using Eq. (8)) for an output wavelength range of 10 nm. It appears to be a straight line whose width is inversely proportional to the length, $\mathcal{L}$ of the waveguide. Under phase-matching condition: $\Delta\beta_{lmn}(\omega_s, \omega_i) = 0$, this can be easily deduced by the argument of the *sinc* function [15]. Now, the *JSA* from Eq. (7) is plotted in Fig. 5(c). It characterizes the joint spectrum of the two photons. From *JSA* we can extract information about the different degrees of freedom of the photon pairs and their quantum correlations. It can be manipulated by the incident pump distribution and phase matching function [16]. Fig. 5(c), shows the negative correlation of the photons pairs. In Fig. 5(d), we plot the *pump envelope intensity: $PEI = |\alpha(\omega_s + \omega_i)|^2$* and in Fig. 5(e) we show the *phase matching intensity: $PMI = |\phi_{lmn}(\omega_s, \omega_i)|^2$*. We notice that the direction of the *PEI* is tilted at $-45°$ (with the positive direction of the horizontal axis) followed from conservation of energy and frequency anti-correlation of the emitted photon pairs. Whereas, the direction of the *PMI*, shown in Fig. 5(e), is at an arbitrary angle $\theta$. This direction can be determined from the gradient of the $\Delta\beta_{lmn}$ and hence is a function of the group velocities of the pump, signal and idler. The gradient of $\Delta\beta_{lmn}$ determines the slope of phase matching intensity [17-19], as follows:

$$\mathrm{grad}\,\Delta\beta_{lmn} = \left(\frac{\partial \Delta\beta_{lmn}}{\partial \omega_s}, \frac{\partial \Delta\beta_{lmn}}{\partial \omega_i}\right)$$

$$= \left(V_{g,p}^{-1}(\omega_p) - V_{g,s}^{-1}(\omega_s),\; V_{g,p}^{-1}(\omega_p) - V_{g,i}^{-1}(\omega_i)\right) \quad (9)$$

Where, $V_{g,\mu}$ is the group velocity defined as:

$$V_{g,\mu} = \frac{d\omega}{dk_\mu(\omega)}, \qquad (\mu = p, s, i) \quad \text{and}$$

In terms of group index:

$$\tan\theta = -\frac{n_p^g(\omega_p) - n_s^g(\omega_s)}{n_p^g(\omega_p) - n_i^g(\omega_i)}, \quad (10)$$

We calculated that the *PMI* in this case is tilted at $\theta \approx -28°$. This can be also seen in Fig. 5(b) and (e).

Now, the Joint Spectral Intensity (*JSI*) is defined as: $JSI = |f_{lmn}((\omega_s, \omega_i)|^2 = PEI \times PMI$. *JSI* corresponds to the probability density of the emitted photon pairs at the frequencies $\omega_s$ and $\omega_i$. The JSI analysis provides an estimation of frequency correlated and frequency decorrelated spectrum. Spectral correlation can be reduced by choosing a PDC process having a PMF with positive slope. The latter is possible only if the value of the pump group index lies between that of the signal and idler. By additionally adjusting the pump bandwidth, we can achieve the uncorrelated spectrum corresponding to spectral separability. On the other hand, with a negative slope of the PMF, there is always complete or some residual frequency correlation. This is shown in Fig. 5(f) for a signal/idler wavelength range of 10 nm. We can easily decipher from Fig. 5(f) that the waveguide is optimally designed for the degenerate emission of frequency correlated photon pairs at 1550 nm.

## 5. COMPARISON OF SIGNAL POWERS GENERATED IN A WAVEGUIDE AND BULK CRYSTAL

We further investigated the spectral density of down-converted signal power for the waveguide (assuming collinear emission for the down conversion process) and compared it with that in bulk crystals. The down-converted signal power in general for waveguides can be obtained from [6] as:

$$dP_s^{(\mathrm{WG})} = \frac{\hbar d^2 P_p \omega_s^2 \omega_i}{\pi \varepsilon_o c^3 n_s n_i n_p} \frac{\mathcal{L}^2}{A_I} \mathrm{sinc}^2\left(\frac{\Delta\beta \mathcal{L}}{2}\right) d\omega_s \quad (11)$$

Where $A_I$ is the effective interaction area. $\omega_{s,i,p}$ are the frequencies of the signal, idler and pump fields, respectively. $\Delta\beta$ is the phase mismatch of the wave vectors.

In case of a bulk crystal, the down-converted signal power can be obtained as:

$$dP_s^{(\mathrm{B})} = \frac{\hbar d^2 \mathcal{L} P_p \omega_s^3 \omega_i^2}{\pi \varepsilon_o c^4 n_p^2 \omega_p} f(\omega_s) d\omega_s \quad (12)$$

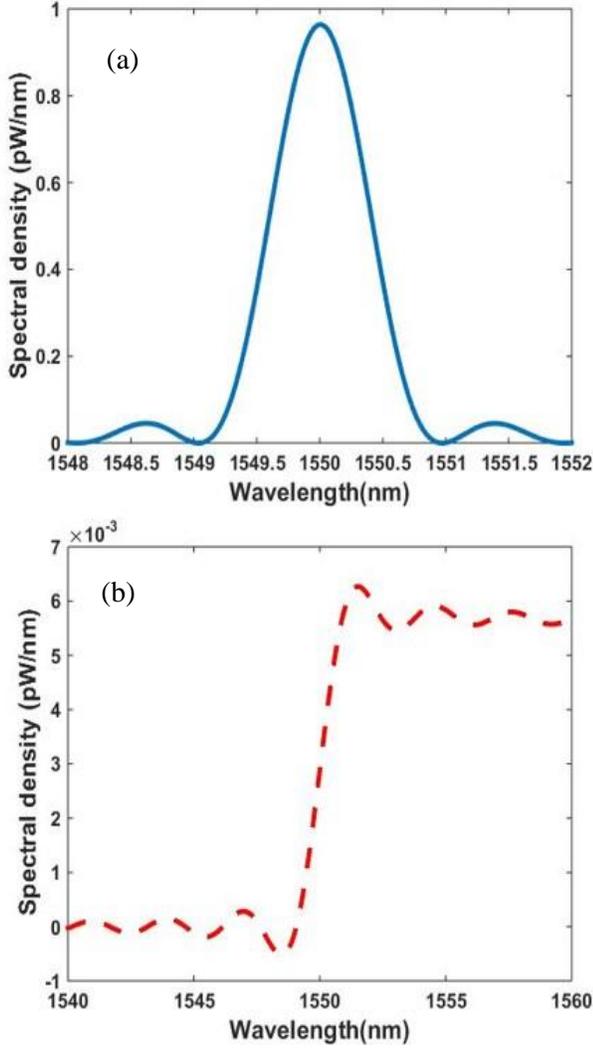

Fig. 6: Signal power densities for (a) waveguide and (b) bulk crystal. The power spectral densities are calculated for a pump power of 1 mW. Notice the different vertical scales for the two plots.

Where $f(\omega_s) = \frac{1}{\pi} \int_0^\infty d(r^2) \, \text{sinc}^2\left[r^2 + \frac{\mathcal{L}}{2}\gamma(\varpi - \omega_s)\right]$ is added for collinear geometry. $\gamma$ is defined as: $\gamma = \partial_{\omega_s} \Delta k|_\varpi - \partial_{\omega_i} \Delta k|_\varpi$, where, $\varpi$ is the degenerate frequency for collinear phase matching. For a perfect collinear emission $\varpi - \omega_s = 0$ when $f(\omega_s) \sim \frac{1.57}{\pi}$. From Eq. (12), we note that the signal power for bulk linearly depends on the pump power as well as the length of the crystal. In Fig. 6, we have plotted the above signal power densities for (a) a waveguide and (b) a bulk crystal. We have taken $A_I = 14.27$ (μm)$^2$, as calculated from the waveguide eigen modes, $d = \frac{2}{\pi} d_{31} = 2.7$ pm/V, $n_p = 2.25$, $n_s = 2.20$, $n_i = 2.12$. We assumed type II degenerate SPDC in ppLN with a 775 nm pump. The main difference between the waveguide and the bulk approach is due to the following. In a bulk medium the pump Gaussian mode interacts with a continuum of plane wave modes, whereas, in waveguides effectively only three modes interact. This alters both the overlap integrals and the density of states. The waveguide emission is confined to a limited band due to the *sinc* function in Eq. (11) whereas Eq. (12) predicts that far from collinear emission bulk spectral density is almost flat.

Let us consider $\omega_s = \frac{\omega_p}{2} + \Omega$ and $\omega_i = \frac{\omega_p}{2} - \Omega$, where $\Omega$ is the frequency detuning from the central frequency $\frac{\omega_p}{2}$ for degenerate operations [7]. For simplicity, let us assume $D = \frac{\hbar d^2 P_p}{\pi \varepsilon_o c^3 n_s n_i n_p} \frac{\mathcal{L}^2}{A_I}$, and using these, Eq. (11) can be rewritten as:

$$dP_s^{(\text{WG})} = D \frac{\omega_p^3}{8} \left(1 + \frac{2\Omega}{\omega_p}\right) \times \left(1 - \left(\frac{2\Omega}{\omega_p}\right)^2\right) \text{sinc}^2\left(\frac{\Delta k \mathcal{L}}{2}\right) d\Omega. \quad (13)$$

We assume the phase matching condition
$$\Delta \beta = \beta_s\left(\frac{\omega_p}{2} + \Omega\right) + \beta_i\left(\frac{\omega_p}{2} - \Omega\right) - \beta_p(\omega_p). \quad (14)$$

A Taylor expansion to quadratic order about the phase matching
$$\beta(\omega) = \beta(\omega_o) + \frac{1}{v}(\omega - \omega_o) + \frac{GVD(\omega_o)}{2}(\omega - \omega_o)^2. \quad (15)$$

Where $v$ is the group velocity and $GVD$ is the group velocity dispersion. Eq. (15) then simplifies to
$$\Delta k = GVD\left(\frac{\omega_p}{2}\right)\Omega^2. \quad (16)$$

We assume that the phase matching condition is satisfied over a range that is small compared to the range of integration. Eq. (13) can be integrated to calculate the down-converted signal power in case of a waveguide (without filters) as:

$$P_s^{(\text{WG})} = \frac{\hbar d^2 P_p \omega_p^3}{6\sqrt{2\pi |GVD(\frac{\omega_p}{2})|}\varepsilon_o c^3 n_s n_i n_p} \frac{\mathcal{L}^{3/2}}{A_I} \quad (17)$$

Note that here we find that the down-converted signal power is directly proportional to $\mathcal{L}^{3/2}$ and inversely proportional to the effective interaction area $A_I$. Hence, field confinement in case of waveguides effectively enhances the overall signal power generated. No such dependence is expected in case of bulk crystals.

For a comparison with bulk systems, we followed the same procedure as above and integrate Eq. (12) for collinear phase-matching to obtain the signal power (to a constant integration factor $C \approx 0.0109$) generated in case of bulk crystals as:

$$P_s^{(\text{bulk})} \approx C \frac{\hbar d^2 \mathcal{L} P_p \omega_p^5}{\pi c^4 \varepsilon_o n_p^2} \quad (18)$$

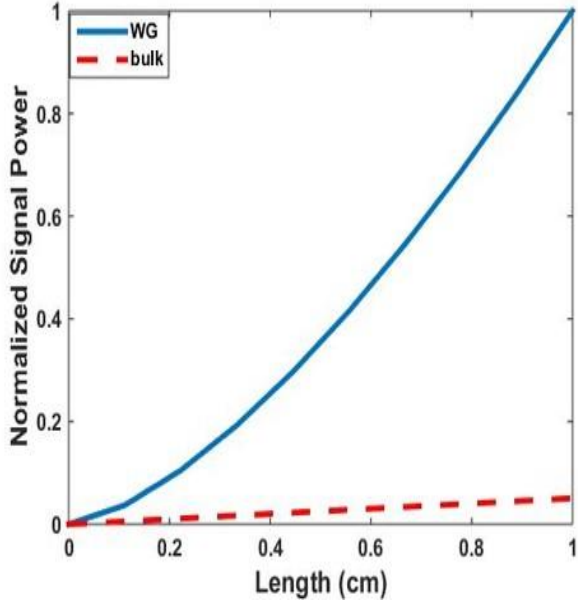

Figure. 7: Calculated signal power as a function of length in waveguides (continuous) and in bulk crystal (dashed).

From Eq. (17) and (18) we find that, in bulk crystals the down-converted signal power is proportional to the length $\mathcal{L}$ of the crystal, whereas in waveguides, it is proportional to $\mathcal{L}^{3/2}$. This dependence is showed in Fig. 7 in case of LN for a pump power of $P_p = 1$ mW, where we have plotted the signal power versus $\mathcal{L}$, considering the same parameters as mentioned above. Here we considered $GVD\left(\frac{\omega_p}{2}\right) = 0.11136$ (ps)$^2$/m at 1550 nm.

## 6. DISCUSSION AND CONCLUSION

The waveguide architecture plays a crucial role for the generation of degenerate photon pairs in desired spatial modes. In particular, ridge waveguides have an advantage over rectangular waveguides in offering an increased bandwidth of operation and a better confinement. The waveguide can be designed in such a way that most higher order spatial modes are suppressed in order to actively enhance the fundamental mode emission for degenerate photon pairs. In addition to using a pump beam with a Gaussian profile, it is also possible to enhance the fundamental mode emission (signal/idler) by reducing the size of the waveguide cross-section. In this aspect nano-waveguides are promising. We have studied the propagation characteristics of a ppLN ridge waveguide structure that will be used for the emission of degenerate photon pairs at 1550 nm. The modal analysis of the SPDC process was done by finite element method (FEMSIM) for the customized ppLN structure to predict the optimal input parameters and also to calculate the effective indices of the pump, signal and idler modes for this waveguide.

We have studied the expected *JSA* and *JSI* of the generated photons pairs considered in the fundamental mode emission and our results predict the emission of *degenerate* photon pairs at 1550 nm. By controlling the *JSI*, we can control the emission properties of the photons pairs, which can be used in many applications. Moreover, the *JSI* analysis provides an estimation of the pump bandwidth that would correspond to a maximum purity of the generated biphotonic state, corresponding to separable states.

Also, since our waveguide supports multiple modes (due to a large cross-section and high index contrast), hence, a complete modal analysis of the structure requires a more detail characterization of the other higher order SPDC processes possible.

Our paper also studies the PDC signal power through an integration of the power spectral density in case of a waveguide and showed its dependence on the length of the waveguide to be $\mathcal{L}^{3/2}$. The result is compared with that of a bulk crystal. Clearly, a higher signal power is obtained in case of a waveguide compared to a same length of a bulk crystal. This can be attributed due to the fact that the waveguide down-conversion is effectively a one-dimensional problem (with the three modes confined) whereas for a bulk it should be treated like a three-dimensional problem. Due to this, the effect is similar to a spectral redistribution of the photon pairs generated in a waveguide and confinement causes an increase of the field overlap, which ultimately enhances the overall efficiency of the SPDC process. Thus it is apparent that waveguide emission is more brilliant than the bulk.

**Acknowledgment**. We thank the Department of Physics, IIT Delhi for the initial seed grant for the purchase of the waveguide.

### REFERENCES

[1] Kwiat P G, Mattle K, Weinfurter H, Zeilinger A, Sergienko A V, Shih Y 1995 *Phys. Rev. Lett*. **75** 4337;
Kwiat P.G, Waks E, White A G, Appelbaum I, Eberhard P H 1999 *Phys. Rev*. A. **60** R773.

[2] Chen J, Pearlman A J, Ling A, Fan J, and Migdall A L 2009 *Opt. Exp*. **17** 6727.
[3] Tanzilli S, Riedmatten H D, Tittel H, Zbinden H, Baldi P, Micheli M D, Ostrowsky D, and Gisin N 2001 *Electron. Lett.* **37** 26.
[4] Banaszek K, U'Ren A. B, and Walmsley I. A 2001 *Opt. Lett*. **26** 1367.
[5] Bharadwaj D and Thyagarajan K 2016 *Phys. Rev*. A **94** 063816.
[6] Fiorentino M, Spillane S M, Beausoleil R G, Roberts T D, Battle P and Munro M W et al. 2007 *Opt. Exp*. **15** 7479.
[7] Helt L G, Liscidini M, and Sipe J E 2012 *J. Opt. Soc. of Am*. B **29** 2199.
[8] Tanzilli S, Tittel W, De Riedmatten H, Zbinden H, Baldi P, De Micheli M, Ostrowsky D B, and Gisin N 2002 *Eur. Phys*. J. D **18** 155.
[9] UmekiT. Tadanaga O and Asobe M., 2010 IEEE J. Quantum Electron. 46, 1206 .
[10] Mohamedelhassan A 2012 M.Sc. Thesis, Department of applied Physics, School of engineering science, KTH, Stockholm, Sweden.
[11] Zhou Y F, Wang L, Liu P, Liu T, Zhang L, Huang D T and Wang X L 2014 *Nuclear Instruments and Methods in Physics Research* B **326,** 110–112.
[12] Zelmon D E, Small D L and Jundt D 1997 *J. Opt. Soc. Am*. B **14** 3319.
[13] Gayer O, Sacks Z, Galun E, Ariel A 2008 *Appl. Phys*. B **91** 343–348.
[14] Christ A, Laiho K, Eckstein A, Lauckner T, Mosley P J and Silberhorn C 2009 *Phys. Rev*. A **80** 033829.
[15] Lemieux S 2016 M.Sc. Thesis, Department of Physics, University of Ottawa.
[16] Dosseva A, Cincio Ł, and Brańczyk A M 2016 *Phys. Rev*. A. **93** 013801.
[17] Jin R B, Shimizu R, Wakui K, Benichi H, and Sasaki M 2013 *Opt. Exp*. **21** 10659.
[18] Laiho K, Pressl B, Schlager A, Suchomel H, Kamp M, Höfling S, Schneider C and Weihs G 2016 *IOP publishing, Nanotechnology* **27** 434003 (7pp).
[19] Eckstein A, 2012 Ph.D thesis, Naturwissenschaftliche Fakult*ä*t der Friedrich-Alexander-Universit*ä*t Erlangen-N*ü*rnberg.